# REGULARITIES AND FEATURES IN THE BEHAVIOR OF ELECTRICAL AND MAGNETIC PROPERTIES OF $Co_2FeZ$ (Z = Al, Si, Ga, Ge, Sn, Sb) HALF-METALLIC FERROMAGNETIC HEUSLER ALLOYS

Yu. A. Perevozchikova*, V. Yu. Irkhin*, A. A. Semiannikova*, V. V. Marchenkov*

*M.N. Mikheev lnstitute of Metal Physics of Ural Branch of Russian Academy of Sciences, 620108 Ekaterinburg, Russia*

*e-mail: march@imp.uran.ru; yu.perevozchikova@imp.uran.ru*

The electrical resistivity, magnetization, and Hall effect in $Co_2FeZ$ (Z = Al, Si, Ga, Ge, Sn, Sb) ferromagnetic Heusler alloys are studied. A number of correlations between the electronic and magnetic characteristics of the studied alloys are revealed while the atomic number of the Z-component changes. For the $Co_2FeAl$ and $Co_2FeSi$ half-metallic ferromagnets, the magnitude of magnetization is consistent with the Slater-Pauling rule. For the $Co_2FeAl$, $Co_2FeSi$, and $Co_2FeGe$ compounds, a quadratic temperature dependence of electrical resistivity at temperatures below 30 K and above 65 K is found. At the same time, in the region of intermediate temperatures (from 40 K to 65 K), a power-law dependence of $\sim T^b$ with an exponent of $3.5 \leq b \leq 4$ is observed, which may be due to two-magnon scattering processes.

Keywords: Heusler alloy, half-metallic ferromagnet, electronic transport, magnetization, Hall effect, electrical resistivity

## 1. INTRODUCTION

Half-metallic ferromagnets (HMFs) are currently promising materials for spintronics, in particular for use in magneto-optical information recording or as spin injectors [1–4]. It has been predicted [1–4] that in the ground state at $T = 0$ K, almost 100% charge carrier spin polarization can be realized due to their unusual band structure (Fig. 1). At the Fermi level for one spin direction (against the magnetization direction, spin down) an energy gap is observed, while there is no gap for the other spin direction. The HMF state can be realized in a wide range of Heusler compounds based on Co, Mn, Fe, etc. For example, in the $Fe_2MeZ$, $Co_2MeZ$, or $Mn_2MeZ$ series, where Me are transition 3$d$-elements (Ti, V, Cr, etc.), and Z are $p$-elements of the periodic table [see, e.g., 5–8]. According to theoretical calculations [8–10], the $Co_2FeAl$, $Co_2FeSi$, and $Co_2FeGe$ compounds are HMFs with complete charge carrier polarization at the Fermi level.

Some experimental and theoretical studies [5–8] have shown that electronic transport and magnetic properties of Heusler alloys vary significantly depending on the density of states at the Fermi level $E_F$ when the composition changing due to the 3$d$-elements. Therefore, to follow the behavior in the electronic and magnetic characteristics of Heusler alloys with changes in $p$-elements, i.e., Z-components, is of great interest.

The $Co_2FeZ$ (Z = Al, Si, Ga, Ge, Sn, Sb) alloys are chosen as study objects. The $Co_2FeAl$, $Co_2FeSi$, and $Co_2FeGe$ compounds included in this group are HMFs with high spin polarization [8–10], and the Curie temperature of these alloys is much higher than room temperature (about 1000 K). Thus, high spin polarization in these compounds can be kept at room temperature.

The electrical resistivity, magnetization, and Hall effect of $Co_2FeZ$ (Z = Al, Si, Ga, Ge, Sn, Sb) Heusler alloys are studied in this work in order to establish the regularities of electronic and magnetic characteristic behavior when the atomic number of the Z-component changes, i.e., when varying $p$-elements.

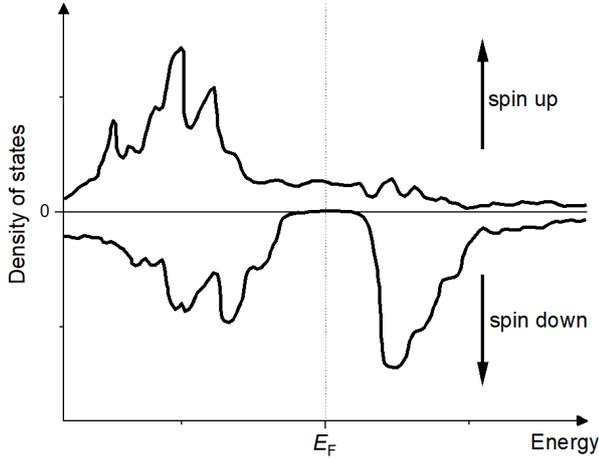

Fig.1. Schematic representation of the density of states in a half-metallic ferromagnet. The arrows indicate the directions of spins for electronic states, the dotted line indicates the Fermi level $E_F$.

## 2. EXPERIMENTAL

The $Co_2FeZ$ ($Z$ = Al, Si, Ga, Ge, Sn, Sb) Heusler alloys were synthesized by arc melting in a purified argon atmosphere, followed by annealing at 773 K for 120 h in a purified argon atmosphere and cooling to room temperature at a rate of ~100 K/hour, as well as additional annealing at 873 K (1073 K for $Co_2FeGa$, $Co_2FeSn$) for 7 days, followed by quenching in ice water to homogenize and obtain the required phase. The atomic content of elements in the polycrystalline alloy was proved using a FEI Company Quanta 200 scanning electron microscope equipped with an EDAX X-ray microanalysis device. The results of the analysis are presented in Table 1. The deviation from the stoichiometric composition in all samples turned out to be insignificant. Structural test showed that in all the studied alloys the main phase is the $L2_1$ structure.

Magnetization was measured using an MPMS-XL-5 SQUID magnetometer by Quantum Design. Electrical resistivity and the Hall effect were measured using a standard four-probe DC method with switching of the electric current flowing through the sample. The structure and magnetic property studies were carried out at the Collaborative Access Center "Testing Center of Nanotechnology and Advanced Materials" of the IMP UB RAS.

**Table 1.** Results of elemental analysis of the samples

| Compound | Composition according to EDAX |
|---|---|
| $Co_2FeAl$ | $Co_{2.04}Fe_{1.07}Al_{0.89}$ |
| $Co_2FeSi$ | $Co_{1.97}Fe_{0.98}Si_{1.05}$ |
| $Co_2FeGa$ | $Co_{1.86}Fe_{1.13}Ga_{1.01}$ |
| $Co_2FeGe$ | $Co_{2.15}Fe_{1.09}Ge_{0.76}$ |
| $Co_2FeSn$ | $Co_{1.90}Fe_{1.04}Sn_{1.06}$ |
| $Co_2FeSb$ | $Co_{1.98}Fe_{1.04}Sb_{0.98}$ |

## 3. RESULTS AND DISCUSSION

### 3.1 Electrical resistivity

The temperature dependences of the electrical resistivity $\rho(T)$ of the studied alloys are presented in Fig. 2. It can be seen that $\rho(T)$ of all alloys increases monotonically with rising temperature, i.e., has a metallic type. The residual resistivity varies from ~ 9 μΩ·cm for $Co_2FeGa$ and $Co_2FeSb$ to ~ 43 μΩ·cm for $Co_2FeAl$ (Table 2).

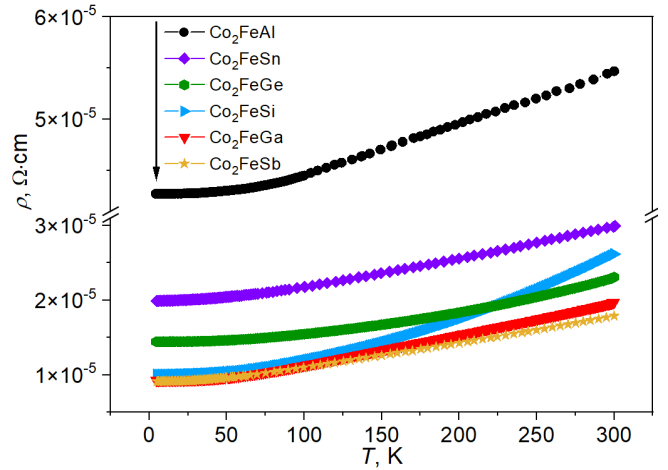

Fig.2. The temperature dependences of the electrical resistivity of the $Co_2FeZ$ ($Z$ = Al, Si, Ga, Ge, Sn, Sb) alloys.

According to theoretical concepts [11], one magnon scattering processes is suppressed and two-magnetic scattering processes may appear in the HMF state, leading to a power-law dependence of the electrical resistivity on temperature. $\rho_{t-m} \sim T^b$, $7/2 < b < 9/2$. Similar contributions to the electrical resistivity have been found experimentally in [12, 13]. On the other hand, in a $Co_2FeSi$ single crystal in the HMF state, a temperature dependence of $\rho(T)$ close to quadratic at low temperatures was observed [14]. Therefore, it is of interest to analyze the form of the $\rho(T)$ dependences in the alloys under study.

**Table 2.** Element atomic number *z*, residual electrical resistivity $r_0$, experimental and reference (calculated or experimental) magnetic moment value corresponding to the saturation magnetization $M_{S\,exp}$ and $M_{S\,ref}$, magnetic moment calculated by the Slater-Pauling rule $M_{Sl-Pol}$, Curie temperature $T_C$ for the $Co_2FeZ$ (Z = Al, Si, Ga, Ge, Sn, Sb) alloys

| Compound | At. No. z (Al, Si, Ga, Ge, Sn, Sb) | $r_0$, μΩ·cm | $M_{S\,exp}$, μB/f.u., T = 4.2 K | $M_{Sl-Pol}$, μB/f.u. | $M_{S\,ref}$, μB/f.u. | $T_C$, K |
|---|---|---|---|---|---|---|
| $Co_2FeAl$ | 13 | 42.7 | 5.4 | 5 | experiment.: 5.27 [15] theor.: 4.99 [16] 4.86-5.22 [17] | experiment.: 1098 [15] 1000 [17] ≥ 1100 [18] theor.: 575-1275 [17] |
| $Co_2FeSi$ | 14 | 10.02 | 5.8 | 6 | experiment.: 5.87 [15] 5.97 [19] theor.: 5.09-5.98 [17] 5.75 [19] | experiment.: 1039 [15] 1100 [17] 1030 [18] 1100±20 [19] theor.: 650-1450 [17] |
| $Co_2FeGa$ | 31 | 9.25 | 6.4 | 5 | experiment.: 5.17 [15] 3.25-5.4 [20] theor.: 5.02 [16] 4.93-5.4 [17] 5 [21] 5.06 [22] | experiment.: 1093 [15] 1056 [18] 1080 [23] theor.: 550-1225 [17] 1252-1369 [19] 1190-1330 [24] |
| $Co_2FeGe$ | 32 | 14.43 | 7.3 | 6 | experiment.: 5.74 (rapid melt quenching) [26] theor.: 5.29-5.98 [17] 5.39; 5.7; 5.99 [25] 5.61 [26] | experiment.: 1060 [18] 1000 [23] 981 (rapid melt quenching) [26] theor.: 475-1350 [17] 972-1141 [19] |
| $Co_2FeSn$ | 50 | 19.87 | 5.1 | 6 | experiment.: 4.3 (82 emu/g at RT, nanoparticles) [27] | experiment.: 968 [18] |
| $Co_2FeSb$ | 51 | 9.13 | 5.4 | 7 | --- | experiment.: ≥ 1100 [18] |

Figure 3 shows the results of the temperature dependence analysis of electrical resistivity from 4.2 K to 75 K. At low temperatures, a quadratic temperature dependence is observed for all compounds, which can be due to both one-magnon and electron-electron scattering. At the same time, for the $Co_2FeAl$, $Co_2FeSi$, and $Co_2FeGe$ alloys at temperatures above 30 K, strong deviations from the $T^2$ law are revealed, which can be described by power-law dependences with higher *b* values. Such deviations are almost absent in $Co_2FeSn$ and $Co_2FeSb$. To determine the exponent at higher temperatures, the corresponding dependences in logarithmic coordinates $\log(\rho-\rho_0-aT^2) = f(\log T)$ were plotted for the alloys $Co_2FeAl$, $Co_2FeSi$, and $Co_2FeGe$. Here $\rho_0$ is the residual electrical resistivity, the coefficient *a* at $T^2$ was determined from experiment in the range from 4.2 to 30 K. Fig. 4 shows that $b > 2$. In the temperature range from 40 to 65 K, for $Co_2FeAl$, $b = 3.5$; for $Co_2FeSi$, $b = 4$; for $Co_2FeGe$, $b = 3.8$. It was in this temperature range that a contribution to the resistivity proportional to $T^b$, where $b = 4$, has been observed in a $Co_2FeSi$ single crystal [13]. This may be one of the manifestations of two-magnon scattering processes predicted in [11]. Recently, a similar behavior has been found in the $Co_2MnGe$ HMF alloy [28].

No power-law dependences of electrical resistivity on temperature with exponent $b > 2$ in the $Co_2FeGa$, $Co_2FeSn$, and $Co_2FeSb$ alloys were observed. In the $Co_2FeGa$ alloy, a transition from a "low temperature" (up to ~25 K) to a "high temperature" (above ~37 K) quadratic temperature dependence of electrical resistivity occurs. According to calculations in [19], $Co_2FeGa$ is close to the state of a half-metallic ferromagnet, however, according to [29], the electrical resistivity is well described by the $T^{2.1}$ law, even though the separation of the highest temperature terms in it was not carried out. Therefore, this situation requires further research.

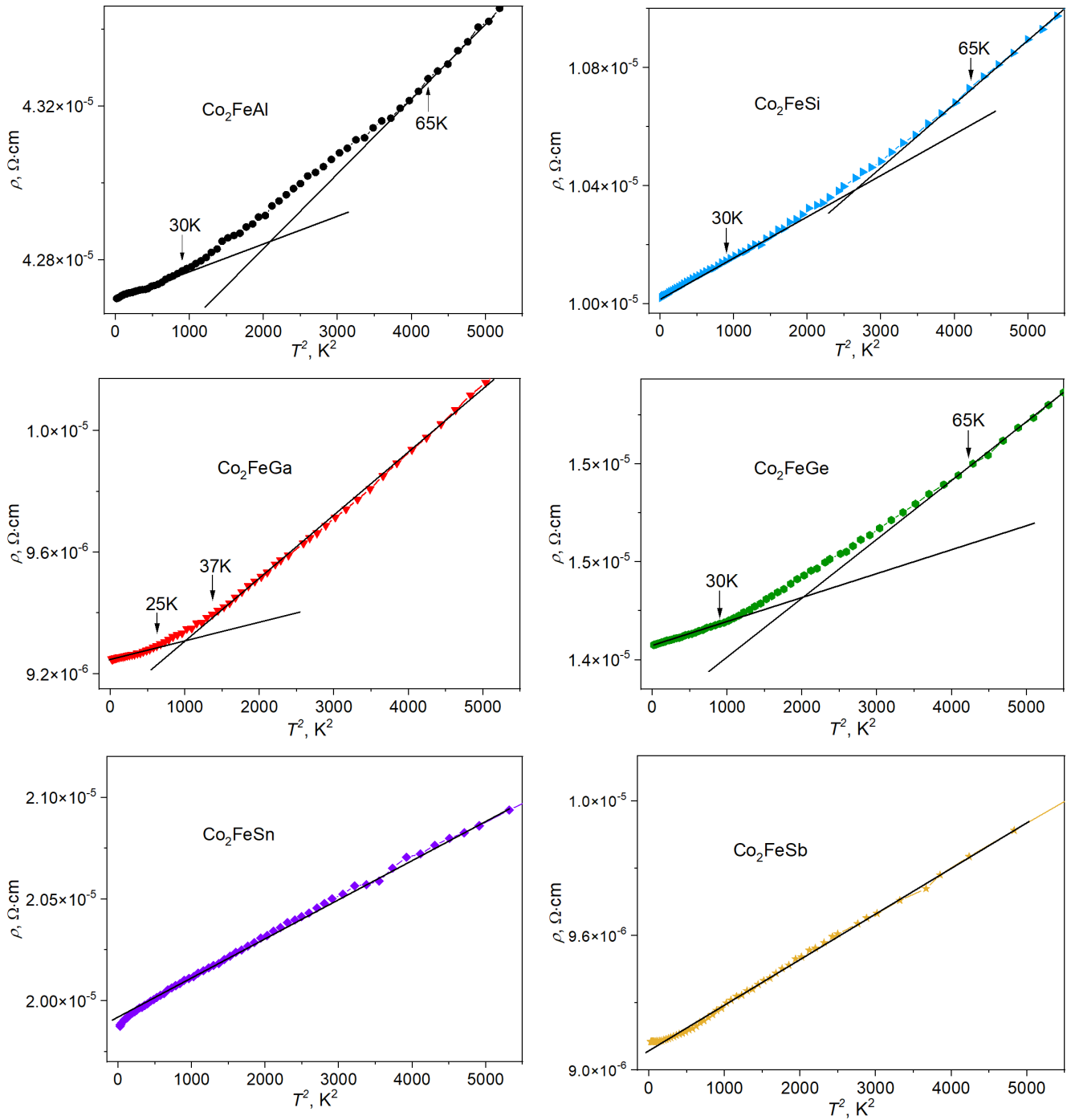

Fig.3. Electrical resistivity dependences of $Co_2FeZ$ ($Z$ = Al, Si, Ga, Ge, Sn, Sb) alloys on the temperature square in the range from 4.2 K to 75 K.

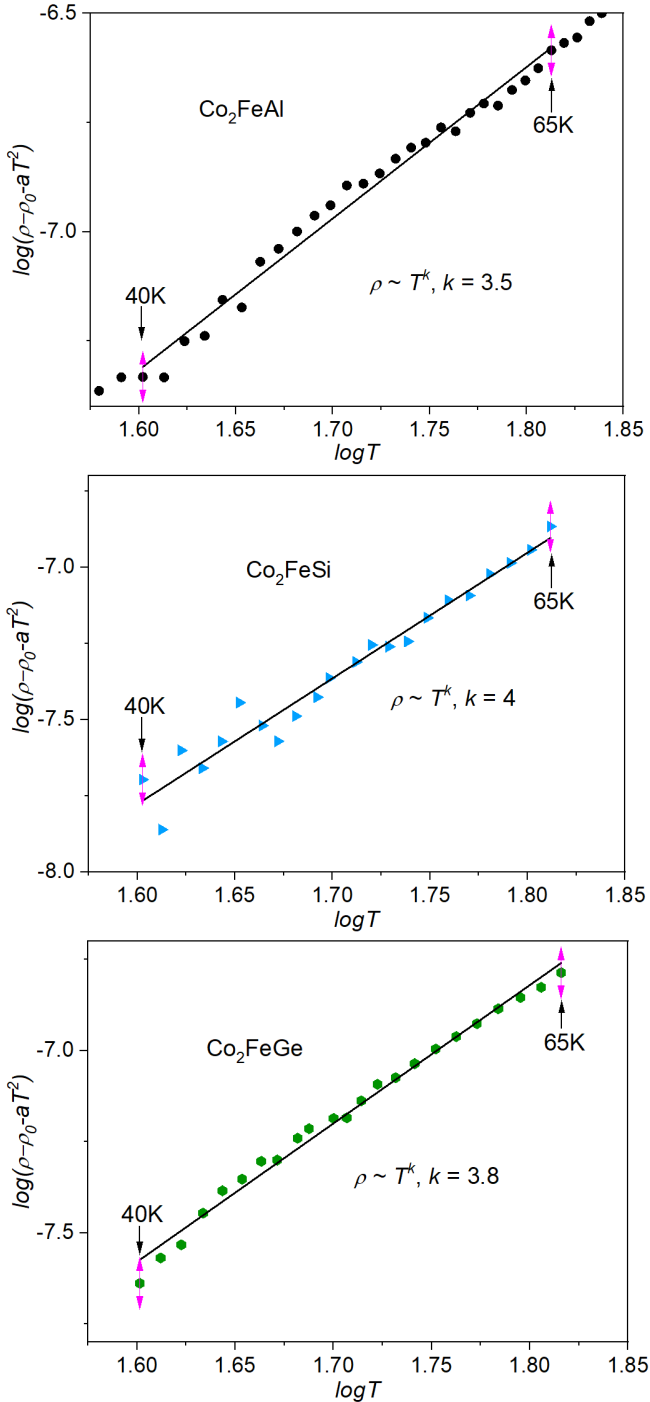

Fig.4. Logarithm $\log(\rho-\rho_0-aT^2)$ dependences on the temperature logarithm $\log(T)$ for $Co_2FeAl$, $Co_2FeSi$, and $Co_2FeGe$ at the temperature range from 40 K to 65 K.

### 3.2 Magnetization

The field dependences of the magnetization of alloys under study at $T = 4.2$ K are shown in Fig. 5. The studied alloys are ferromagnetic up to $T_C$, and in fields above 20 kOe their magnetization reaches saturation. The saturation magnetization $M_{S\,exp}$ values, which were determined as the $M$ values in a field of 50 kOe at $T = 4.2$ K, are presented in Table 2. For comparison, the table presents data on saturation magnetization $M_{S\,ref}$ from [15–27]. According to the Slater-Pauling rule, the total spin magnetic moment $M_t$ is related to the total number of valence electrons $Z_t$ by a simple expression: $M_t = Z_t–24$ [30]. Therefore, the magnetic moment for the $Co_2FeAl$, $Co_2FeGa$ alloys should be equal to 5 $\mu_B$/f.u., 6 $\mu_B$/f.u. value is expected for the $Co_2FeSi$, $Co_2FeGe$, $Co_2FeSn$, and 7 $\mu_B$/f.u. magnetic moment is predicted for the $Co_2FeSb$ (Table 2). However, the saturation magnetization values are close to theoretical only for $Co_2FeAl$ and $Co_2FeSi$. The experimentally determined magnetization is presumably inconsistent with predictions in [30] due to deviations from stoichiometry and inhomogeneities in the compositions of the $Co_2FeGa$ and $Co_2FeGe$ samples. According to [29], for $Co_2FeGa$ the magnetization approximately equals to ≈ 120 emu/g, which corresponds to a magnetic moment of ~5.2 $\mu_B$/f.u.

### 3.3 Hall effect

Figure 6 shows the Hall resistivity $\rho_H$ dependence on the magnetic field. Using these dependences the coefficients of normal and anomalous Hall effects are determined, as well as the type of main charge carriers: holes for $Co_2FeSi$ and $Co_2FeGe$, electrons for the others. The concentration and mobility of main charge carriers were estimated using a one-band model and the method described in [31]. The results are presented in Table 3. It is worth noting that the measurements were performed on polycrystalline samples, therefore, the assessments of the concentration and mobility of charge carriers are qualitative. Nevertheless, even such qualitative assessments make it possible to trace regularities in the electronic characteristics with a change in the $Z$-component in $X_2YZ$ Heusler alloys (see, for example, [5, 6, 18]).

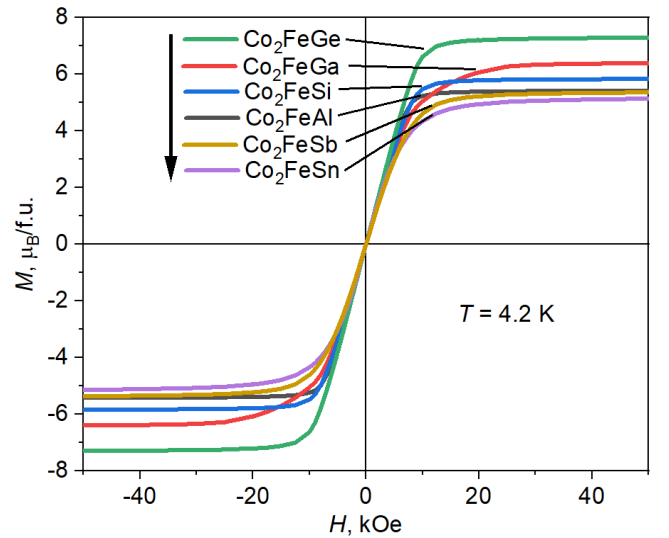

Fig.5. Field dependences of the magnetization of the studied alloys at $T$= 4.2 K.

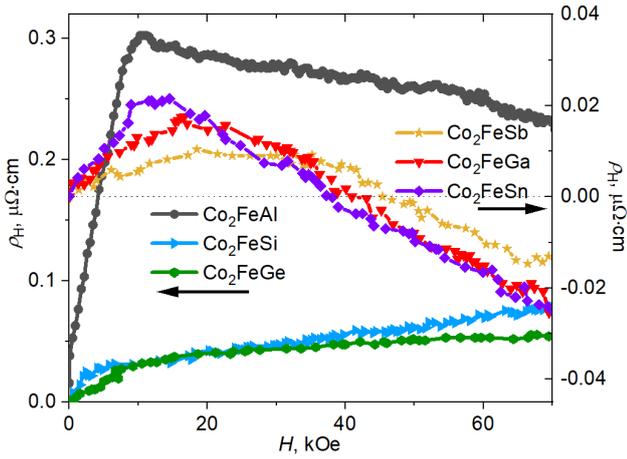 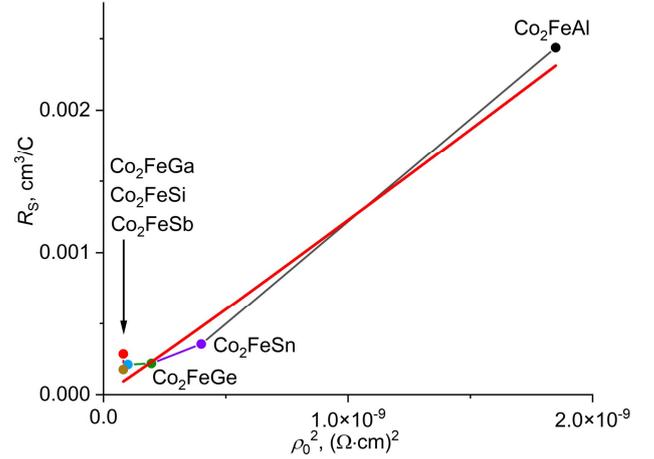

Fig.6. Field dependences of the Hall resistivity of the studied alloys at $T = 4.2$ K.

Fig.7. Dependence of the of the anomalous Hall coefficient $R_S$ on the square of the residual electrical resistivity $\rho_0^2$.

According to the theory of the anomalous Hall effect [31], the anomalous Hall coefficient $R_S$ contains contributions from various scattering mechanisms. Scattering my occur on impurities, phonons, and spin inhomogeneities, which lead to both linear and quadratic terms in the $R_S$ dependence on the electrical resistivity $\rho$. Fig. 7 shows that this dependence for the studied Co$_2$FeZ (Z = Al, Si, Ga, Ge, Sn, Sb) alloys can be described as $R_S \sim \rho^k$, where $k = 2.06 \pm 0.18$.

3.4 Dependences of electronic and magnetic characteristics on the atomic number of the Z-component

When the Z component in the Co$_2$FeZ (Z = Al, Si, Ga, Ge, Sn, Sb) alloys varies, their magnetic and electronic properties change. To visualize these regularities, Fig. 8 demonstrates the dependences of the residual electrical resistivity, saturation magnetization, and normal and anomalous Hall effect coefficients on the atomic number z of Z-component.

The residual electrical resistivity is maximum for Co$_2$FeAl. With increasing atomic number, it decreases for Co$_2$FeSi and Co$_2$FeGa, then rises again, reaching a local maximum for Co$_2$FeSn, and decreases for Co$_2$FeSb (Fig. 8a).

In the simplest one-band model, the conductivity $\sigma$ (i.e., the inverse electrical resistivity $1/\rho$) is proportional to the concentration of charge carriers $n$ and their mobility $\mu$. Using the experimentally determined values of the concentration and mobility of current carriers (Table 3), the dependences $(\mu \cdot n)^{-1} = f(z)$ were plotted for all studied alloys (Fig. 8b). It is evident that the behavior of $\rho$ is qualitatively well matched to $(\mu \cdot n)^{-1}$ with a change in the atomic number of the Z-component of Co$_2$FeZ alloys.

Certain correlations are observed between the electrical resistivity and both the Hall coefficients, as well as between the latter. The maximum value of electrical resistivity and anomalous Hall coefficient occurs for Co$_2$FeAl (Fig. 8a and 8c). The local maximum of electrical resistivity coincides with that for the anomalous Hall in the Co$_2$FeSn alloy. With a further increase in the atomic number, when moving to Co$_2$FeSb, a decrease in both the electrical resistivity and the anomalous Hall coefficient is observed (Fig. 8a and 8c). From a comparison of the anomalous (Fig. 8c) and normal (Fig. 8d) Hall coefficients, it is clear that the maximum values of $R_S$ correspond to the minimum values of $R_0$, and vice versa.

**Table 3.** Atomic element number z, main charge carriers type, coefficient of the normal Hall effect $R_0$, coefficient of the anomalous Hall effect $R_S$, concentration $n$ and mobility $\mu$ of the charge carriers.

| Compound | At. No. z (Al, Si, Ga, Ge, Sn, Sb) | The type of main charge carriers (electrons or holes) | $R_0$, $10^{-5}$ cm$^3$/C | $R_S$, $10^{-5}$ cm$^3$/C | $n$, $10^{22}$ cm$^{-3}$ | $\mu$, cm$^2$/(V·s) |
|---|---|---|---|---|---|---|
| Co$_2$FeAl | 13 | electrons | -9 | 244 | 7 | 2 |
| Co$_2$FeSi | 14 | holes | 7 | 21 | 9 | 7 |
| Co$_2$FeGa | 31 | electrons | -10 | 29 | 6.5 | 10.6 |
| Co$_2$FeGe | 32 | holes | 4 | 22 | 17 | 2.6 |
| Co$_2$FeSn | 50 | electrons | -9 | 36 | 7 | 4.5 |
| Co$_2$FeSb | 51 | electrons | -3 | 18 | 19 | 3.6 |

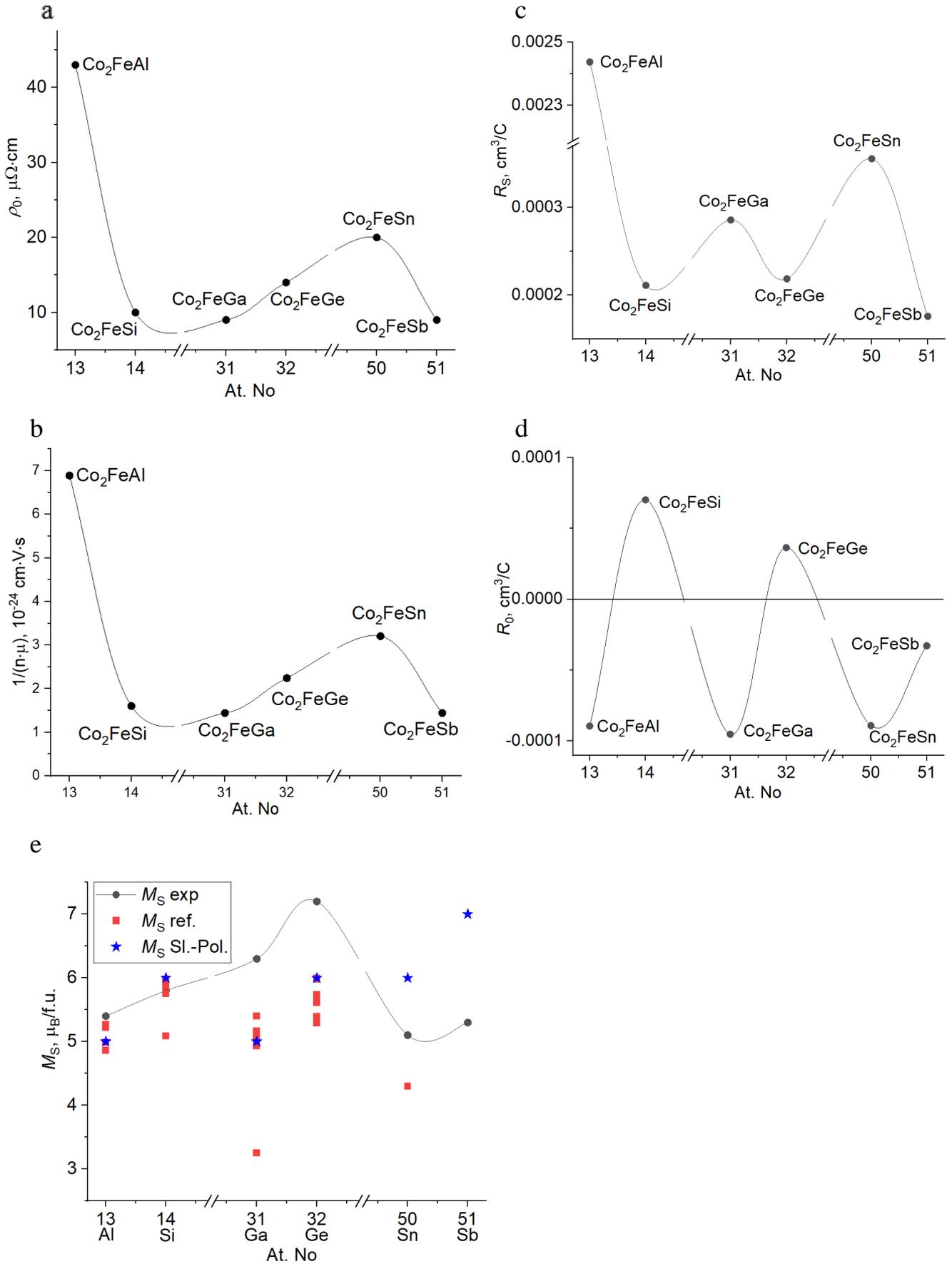

Fig. 8. Dependences of (a) residual electrical resistivity $\rho_0$, (b) the inverse product of concentration by mobility $1/(n \cdot \mu)$, (c) anomalous $R_S$ and (d) normal $R_0$ Hall coefficients, and (e) saturation magnetization $M_S$ on the atomic number of element $Z$.

Figure 8e shows that the values of the magnetic moment obtained in this study, as well as experimental data and theoretical calculations by other authors, differ greatly, with the exception of the magnetic moments for $Co_2FeAl$ and $Co_2FeSi$. The magnetic moment of the alloys under study increases monotonically from 5.4 $\mu_B$/f.u. for $Co_2FeAl$ to 7.2 $\mu_B$/f.u. for $Co_2FeGe$, then decreasing to 5.1 $\mu_B$/f.u. and 5.3 $\mu_B$/f.u. for $Co_2FeSn$ and $Co_2FeSb$, respectively.

The trends of change with atomic number of $Z$-elements of the magnetic moment in the HMF as an integral characteristic are determined by the Slater-Pauling rule. Similarly, in a simple one-band model, the normal Hall coefficient is determined only by the number of charge carriers. At the same time, other kinetic characteristics significantly depend on the scattering process details, therefore, on the spectrum of electronic states near the Fermi level. Thus, the explanation of the corresponding correlations is more complicated. In [18], it was demonstrated that the electrical resistivity with increasing temperature behaves as the square of spontaneous magnetization, depending on the number of charge carriers (in these alloys, the values of $M_{spont}$ are close to the values of $M_{S\,exp}$).

## 4. CONCLUSION

The conduction studies of the magnetic and electrical characteristics of the $Co_2FeZ$ ($Z$ = Al, Si, Ga, Ge, Sn, Sb) Heusler compounds obtained the following results.

For $Co_2FeAl$, $Co_2FeSi$, and $Co_2FeGe$ alloys, a quadratic temperature dependence of electrical resistivity is observed at $T < 30$ K, and in the intermediate temperature region (from 40 K to 65 K) a power-law dependence $\sim T^b$ with an exponent of $3.5 \le b \le 4$ appears. This may indicate two-magnon scattering processes prevailing in HMF materials [11].

The anomalous Hall coefficient $R_S$ is found to be approximately proportional to the square of electrical resistivity $\rho^2$, which is consistent with theoretical concepts [31].

It is shown that the saturation magnetization values are close to the theoretical values only for $Co_2FeAl$ and $Co_2FeSi$. Values for other alloys are substantially different from theoretical ones, however, they agree well with some literature data. The difference between the values obtained in the experiment and theoretical predictions in [30] may be caused by a deviation from stoichiometry and the presence of inhomogeneities in the compositions of $Co_2FeGa$ and $Co_2FeGe$ samples.

A number of correlations between the electronic and magnetic characteristics of the studied alloys is demonstrated, which occur when the atomic number of the $Z$-component changes, i.e., the $p$-elements vary.

The obtained regularities can be used in the selection of materials with optimal characteristics for spintronic devices, when the state of half-metallic ferromagnetism is most favorable. The $Co_2FeSi$ alloy can be singled out among the compounds obtained as promising material for practical use, since a significant value of charge carrier spin polarization has been experimentally observed in it [32, 33] with a relatively low residual electrical resistivity and a sufficiently high magnetic moment.

The work was performed within the framework of the state assignment of the Ministry of Science and Higher Education of the Russian Federation (themes "Spin" No. 122021000036-3 and "Quantum" No. 122021000038-7). The authors thank E.B. Marchenkova for her help in carrying out this work.